\documentclass[prd,a4paper,showpacs,preprint,byrevtex]{revtex4}
\usepackage{graphics}

\begin{document}

\title{Casimir scaling, glueballs and hybrid gluelumps}

\author{Vincent Mathieu}
\thanks{IISN Scientific Research Worker}
\email[E-mail: ]{vincent.mathieu@umh.ac.be}
\author{Claude \surname{Semay}}
\thanks{FNRS Research Associate}
\email[E-mail: ]{claude.semay@umh.ac.be}
\affiliation{Groupe de Physique Nucl\'{e}aire Th\'{e}orique,
Universit\'{e} de Mons-Hainaut,
Acad\'{e}mie universitaire Wallonie-Bruxelles,
Place du Parc 20, BE-7000 Mons, Belgium}
\author{Fabian Brau}
\email[E-mail: ]{f.brau@cwi.nl}
\affiliation{
CWI, P.O. Box 94079, 1090 GB Amsterdam, The Netherlands}

\date{\today}

\begin{abstract}
Assuming that the Casimir scaling hypothesis is well verified in QCD,
masses of glueballs and hybrid gluelumps (gluon with a point-like
$c\bar c$ pair) are computed within the rotating string formalism. In
our model, two gluons are attached by an adjoint string in a glueball
while the gluon and the colour octet $c\bar c$ pair are attached by two
fundamental strings in a hybrid gluelump. Masses for such exotic hadrons
are computed with very few free parameters. These predictions can serve
as a guide for experimental searches. In particular, the ground state
glueballs lie on a Regge trajectory and the lightest $2^{++}$ state has
a mass compatible with some experimental candidates.
\end{abstract}



\maketitle

\section{Introduction}
\label{sec:intro}

Lattice calculations \cite{bali00} and other models of QCD
\cite{hans86,sema04a} predict that the Casimir scaling hypothesis is
well verified in QCD, that is to say that the potential between two
opposite colour charges in a colour singlet is proportional to the value
of the quadratic Casimir operator. This mechanism can be tested for
instance in two-gluon glueballs. Direct evidence for such a system is
still controversial, but several models \cite{corn83,morn99,brau05}
predict similar masses with values close to some of the experimental
candidates \cite{zou99,bugg00}. The Casimir scaling hypothesis can also
be tested in another system whose colour-spin structure is similar to
the one of a glueball. Let us consider an hybrid meson containing a
colour octet spin one $c\bar c$ pair and a gluon. Due their very heavy
masses, the charm quarks can be assumed fixed at the centre of mass with
the gluon orbiting around. A gluon attached to a heavy colour octet is
called a gluelump, so the $gc\bar c$ system considered here is called
hybrid gluelump as in Ref.~\cite{abre05}.

In our model, a quark and an antiquark are attached by a fundamental
string while two gluons in a glueball are attached by an adjoint string.
We assume here that the three-body nature of the confinement implies
that the gluon and the colour octet pair are attached by two fundamental
strings in a hybrid gluelump. Within these hypothesis, we compute the
masses of glueballs and hybrid gluelumps with a simple effective model
of QCD \cite{guba94,morg99}. This semirelativistic potential model is
described in Sec.~\ref{sec:ham}. It depends on few parameters which are
fixed on well known mesons in Sec.~\ref{sec:param}. Masses of glueballs
and hybrid gluelumps are computed respectively in Sec.~\ref{sec:gg} and
\ref{sec:hyb}. We focus our attention on the first orbital excitations,
neglecting the details of the spin dependent interactions, assumed weak
with respect to the confinement. Some concluding remarks are given in
Sec.~\ref{sec:conclu}.

\section{Hamiltonian}
\label{sec:ham}

Starting from the QCD theory, a Lagrangian for a system of two confined
spinless colour sources can be derived taking into account the dynamical
degrees of freedom of the string, with tension $\sigma$, joining the two
particles, the rotating string model (RSM) \cite{guba94,morg99}. This
model is completely equivalent to the relativistic flux tube model
\cite{laco89}, once the auxiliary fields appearing in the RSM are
properly eliminated \cite{alle03,sema04b,buis04}. The nonlinear coupled
equations of these models are difficult to solve \cite{laco89,buis05a}.
So we will use an approximation of these effective QCD theories
\cite{bada02,buis05b}, which is simply given by the following spinless
Salpeter Hamiltonian
\begin{equation}
\label{ham}
H_0 = \sqrt{\vec p^2+m_1^2} + \sqrt{\vec p^2+m_2^2}  + \sigma r,
\end{equation}
completed by a perturbative part due to the motion of the string
\begin{eqnarray}
\label{hstr}
\lefteqn{\Delta H_{\rm str} = -\frac{\sigma L (L+1)}{r}} \nonumber \\
&&\times
\frac{\left[ 4(\mu_1^2 + \mu_2^2 - \mu_1 \mu_2) + (\mu_1 + \mu_2) \sigma
r\right]}{2 \mu_1 \mu_2 \left[ 12 \mu_1 \mu_2 + 4 (\mu_1 + \mu_2) \sigma
r + (\sigma r)^2\right]}.
\end{eqnarray}
The quantity $\mu_i$ appearing in the above equation is a kind of
constituent particle mass given by
\begin{equation}
\label{mu}
\mu_i = \left\langle \sqrt{\vec p^2+m_i^2} \right\rangle,
\end{equation}
in which the average value is taken for an eigenstate of the
Hamiltonian~(\ref{ham}). The constituent mass is then state-dependent.
Equation~(\ref{hstr}) is obtained by taking the slow motion limit of
Eqs.~(33) in Ref.~\cite{buis04}.
The contribution $\Delta M_{\rm str}$ to the mass due to the
string can be computed with a good precision
by the following approximation \cite{bada02,buis05b}
\begin{eqnarray}
\label{mstr}
\lefteqn{\Delta M_{\rm str} = -\sigma L (L+1) \left\langle \frac{1}{r}
\right \rangle} \nonumber \\
&&\times
\frac{\left[ 4(\mu_1^2 + \mu_2^2 - \mu_1 \mu_2) + (\mu_1 + \mu_2) \sigma
\langle r \rangle \right]}{2 \mu_1 \mu_2 \left[ 12 \mu_1 \mu_2 + 4 (
\mu_1 + \mu_2) \sigma \langle r \rangle  + (\sigma \langle r \rangle )^2
\right]}.
\end{eqnarray}
It is worth noting that, in this formalism, the particle masses $m_1$
and $m_2$ are the current ones. So, in the following, we will consider
that the mass $m_n$ of the $n$ quark ($n$ stands for $u$ or $d$) and the
mass $m_g$ of the gluon are vanishing.

A significant contribution to hadron masses is also given by the
one-gluon exchange (OGE) mechanism between colour sources. At the zero
order (neglecting the spin), the interaction has the following form
\begin{equation}
\label{hcoul}
\Delta H_{\rm Coul} = - \kappa \frac{\alpha_S}{r}.
\end{equation}
$\kappa$ is a colour factor given by
\begin{equation}
\label{kappa}
\kappa = \frac{1}{2} \left( C_{12} - C_1 - C_2 \right),
\end{equation}
where $C_{12}$ is the colour Casimir operator for the pair $12$ and
$C_i$ is the corresponding one for the particle $i$.

If $M_0$ is the solution of the eigenequation
$H_0|\phi\rangle = M_0|\phi\rangle$, the total mass $M$ for the system
of the two colour sources is given by
\begin{equation}
\label{m}
M = M_0 + \Delta M_{\rm str} + \langle \Delta H_{\rm Coul} \rangle.
\end{equation}
In this formula, the Coulomb-like contribution is computed as a
perturbation. It is shown in Ref.~\cite{bada02}, that this
approximation is very good, especially for states with high angular
momentum $L$. All computation are performed using the Lagrange-mesh
method \cite{sema01}.

Recently, it was shown that the quark self-energy (QSE) contribution,
which is created by the colour magnetic moment of the quark propagating
through the vacuum background field, adds a negative constant to the
hadron masses \cite{simo01}. Its negative sign is due to the
paramagnetic nature of the particular mechanism at work in this case.
Other contributions due to quark spin (spin-spin, spin-orbit) also exist
but they are smaller that the QSE one \cite{dubi93}, and they are
neglected in this work. The QSE contribution $\Delta S_i$ for a quark of
current mass $m_i$ is given by
\begin{equation}
\label{mqse}
\Delta S_i = - f \frac{\sigma}{2 \pi} \frac{\eta(m_i/\delta)}{\mu_i}.
\end{equation}
The $\eta$ function is such that $\eta(0)=1$, and its value decreases
monotonically towards 0 with increasing quark mass \cite{buis05b}.
$\delta$ is the inverse of the gluonic correlation length, and its value
is estimated about 1.0-1.3~GeV. As the meson masses vary very little
with this parameter \cite{buis05b}, we fix the $\delta$ value at 1~GeV
(meson masses do not depend on $\delta$ when $m=0$).
The factor $f$ has been computed by lattice calculations. First quenched
calculations gave $f=4$ \cite{digi92}. A more recent unquenched work
\cite{digi04} gives $f=3$, the value that we choose in this work. If a
hadron contains
$q$ quarks, a contribution $\sum_{i=1}^q \Delta S_i$ must be added to
its mass. Strong theoretical and phenomenological arguments indicate
that gluons do not bring any contribution of self energy
\cite{kaid00,simo05}.

\section{Parameters}
\label{sec:param}

As we are mainly interested in describing the main features of spectra,
we can use a Hamiltonian of spinless particles since spin
effects are generally an order of magnitude smaller than orbital or
radial excitations. Moreover, we will consider hadrons for which the
spin-dependent part of the Hamiltonian is the weakest possible.

The raise of degeneracy between spin 0 and spin 1 mesons can be due to a
spin-spin potential coming from the OGE or due to an instanton induced
interaction \cite{blas90,brau98}. For spin 1 mesons, the spin-dependent
interaction is small in the former case or vanishing in the latter case.
The spin-dependent part of the gluon-gluon interaction has a small
contribution for glueballs with spin 2 \cite{brau05}. If we consider a
hybrid gluelump containing a $c\bar c$ pair with spin 1, we minimize the
spin effect for the heavy quarks. As such a system is similar to a
glueball from the point of view of the spin-colour structure, we can
also expect that the spin contribution will be the smallest if we study
spin 2 hybrids.
For all these reasons, as in Ref.~\cite{abre05}, we will only consider
systems with maximal value of the total spin $S$: $J=L+1$ for mesons and
$J=L+2$ for glueballs and hybrid gluelumps.

A correct determination of the parameter is crucial to predict new
states. For theoretical reasons, we have chosen to take $m_n=0$,
$m_g=0$, $f=3$ and $\delta=1$~GeV. Now, we will fix the values of the
remaining free parameters with well established meson states. As, the
light $n\bar n$ spin 1 mesons can be isoscalar ($I=0$) or isovector
($I=1$), we will use an average mass defined, as usual, by
\begin{equation}
\label{mav}
M_{n\bar n} ({\rm av.}) = \frac{M_{n\bar n}(I=0) + 3\, M_{n\bar
n}(I=1)}{4}.
\end{equation}

\begin{table}
\caption{Experimental masses in GeV for some light mesons ($n\bar n$),
$D^*$ mesons ($c\bar n$), and charmoniums ($c\bar c$), with maximal $S$
and $J=L+S$ \cite{pdg04}. The mesons $\psi(4040)$ and $\psi(4415)$ are
assigned to $N=2$ and $N=3$ states respectively.}
\label{tab:exp}
\begin{tabular}{lrrrr}
\hline\noalign{\smallskip}
$N=0$, $J^{P(C)}$ & $1^{-(-)}$ & $2^{+(+)}$ & $3^{-(-)}$ &
$4^{+(+)}$ \\
\noalign{\smallskip}\hline\noalign{\smallskip}
$n\bar n$ $I=0$ & 0.783 & 1.275 & 1.667 & 2.034 \\
$n\bar n$ $I=1$ & 0.775 & 1.318 & 1.689 & 2.001 \\
$n\bar n$ av.   & 0.777 & 1.307 & 1.684 & 2.009 \\
$c\bar n$ & 2.010 & 2.460 & & \\
$c\bar c$ & 3.097 & 3.556 & & \\
\noalign{\smallskip}\hline\noalign{\smallskip}
$J=1$, $N$ & 0 & 1 & 2 & 3 \\
\noalign{\smallskip}\hline\noalign{\smallskip}
$c\bar c$ & 3.097 & 3.686 & 4.040 & 4.415 \\
\noalign{\smallskip}\hline
\end{tabular}
\end{table}

At this stage, it is interesting to obtain an approximate analytical
formula for light $n\bar n$ mesons. For such a system, $m_1=m_2=0$ and
$\mu_1=\mu_2=\mu = \left\langle \sqrt{\vec p^2} \right\rangle$. It can
then be shown that $M_0=4\mu$ with $\mu \propto \sqrt{\sigma}$
\cite{sema04b}.
Using the semirelativistic virial theorem \cite{luch90}, we find
$\langle r \rangle = 2 \mu/\sigma$. With the following approximation
$\langle 1/r \rangle \approx 1/\langle r \rangle$, we obtain finally
\begin{equation}
\label{mapprox}
M_{n\bar n} \approx \sqrt{\sigma} \left( 4 \nu - \frac{L(L+1)}{16 \nu^3}
- \frac{\kappa \alpha_S}{2 \nu} - \frac{f}{\pi\nu}\right)
\end{equation}
in which $\nu=\mu/\sqrt{\sigma}$ is independent of $\sigma$.
Let us note that an approximate
value for $\nu$ is given by \cite{buis05b}
\begin{equation}
\label{nuapprox}
\nu \approx \left( \frac{\epsilon_{NL}}{3} \right)^{3/4},
\end{equation}
where $\epsilon_{NL}$ is the solution of the dimensionless Hamiltonian
$\vec q^2 + |\vec x|$ in which $\vec q$ and $\vec x$ are conjugate
variables. The accuracy of formula~(\ref{mapprox}) with the
approximation~(\ref{nuapprox}) is about 5-10\% \cite{bada02}.

The contribution of the motion of the string to the meson mass is small,
since the ratio $\Delta M_{\rm str}/M_0$ is at most 1/16 \cite{buis05b}.
Formula~(\ref{mapprox}) shows that the contribution of the Coulomb term
and of the self energy shift the square mass $M_{n\bar n}^2$ of the
meson without modifying the slope of the Regge trajectories
\cite{buis05b}. This slopes depend mainly of the string tension which we
call $a$ for a quark-antiquark bound state. With a fit, it is only
possible to determine the following combination of parameters
$\kappa \alpha_S/2 -f/\pi$. But since the value of $f$ is fixed, it is
possible to compute $\alpha_S$.

We fix the parameter by fitting the exact results of
Hamiltonian~(\ref{ham}) on experimental data \cite{pdg04}
(see Table~\ref{tab:exp}).
Using the Regge trajectory for light meson, we find $a=0.175$~GeV$^2$
and $\alpha_s=0.10$. The value obtained for $\alpha_S$ is small with
respect to values found in other potential models of meson. For
instance, in Ref.~\cite{bada02}, a value of 0.39 is obtained because the
meson masses are fitted on the centre of gravity of spin 1 and spin 0
mesons. In this case, the average mass of the $J=1$ $n\bar n$ state is
0.612~ GeV, a value quite different from the one we choose, 0.777~GeV.
The mass of the quark $c$ is determined by computing the masses of $D^*$
mesons, which are systems dynamically closer to hybrid gluelumps than
charmonium states. All parameters are gathered in Table~\ref{tab:param}.

\begin{table}
\caption{Parameters of the model. The values of $m_n$, $m_g$,
$m_{c\bar c}$, $f$, and $\delta$  are fixed by theoretical or
phenomenological considerations (see text).}
\label{tab:param}
\begin{tabular}{ll}
\hline\noalign{\smallskip}
$m_n = 0$ & $a=0.175$~GeV$^2$ \\
$m_g = 0$ & $\alpha_s=0.10$ \\
$m_c = 1.300$~GeV & $f=3$ \\
$m_{c\bar c} = 2\, m_c$ & $\delta=1$~GeV \\
\noalign{\smallskip}\hline
\end{tabular}
\end{table}

\begin{figure}
\resizebox{0.50\textwidth}{!}{%
  \includegraphics{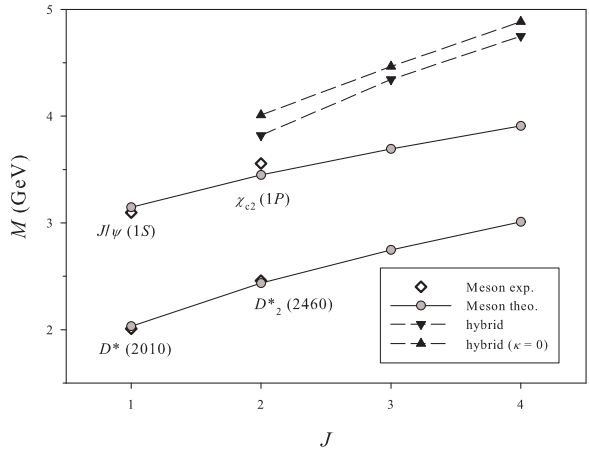}
}
\caption{Masses for $c\bar n$, $c\bar c$, and hybrid gluelumps (with or
without Coulomb interaction). These states are characterized by $N=0$,
$S$ maximal, and $J=L+S$.}
\label{fig:ccg}
\end{figure}

\begin{figure}
\resizebox{0.50\textwidth}{!}{%
  \includegraphics{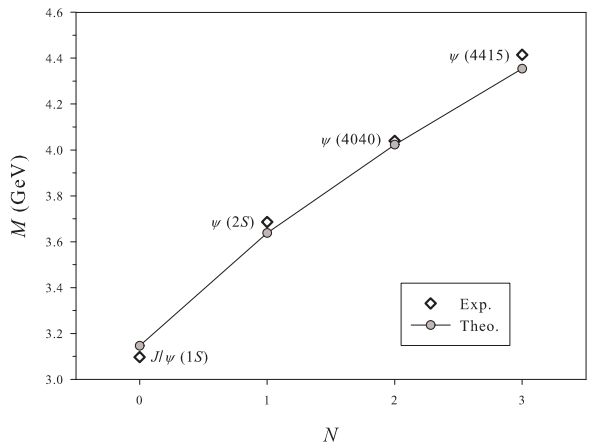}
}
\caption{Masses of $c\bar c$ $J=1$ mesons as a function of the quantum
radial number $N$. The states $\psi(4040)$ and $\psi(4415)$ are assigned
to $\psi(3S)$ and $\psi(4S)$ respectively.}
\label{fig:cc}
\end{figure}

The quality of a fit is estimated by computing for each state the
relative error on binding energy. The theoretical (experimental) binding
energy for a meson is given by the theoretical (experimental) mass minus
the theoretical current masses of the quark and the antiquark (0 for the
light $n\bar n$ mesons, $m_c$ for the $D^*$, $2\,m_c$ for the
charmoniums). Quite good meson masses are obtained, taking into account
the simplicity of the model (see Table~\ref{tab:m}). The error is around
1\% for light $n\bar n$ mesons (see Fig.~\ref{fig:gg}), around 3\% for
the $D^*$ (see Fig.~\ref{fig:ccg}); around 10\% for the mesons $J/\psi$
and $\chi_{c2}$ (see Fig.~\ref{fig:ccg}), but less than 5\% for radial
excitations of the $J/\psi$ (see Fig.~\ref{fig:cc}).

\section{Glueballs}
\label{sec:gg}

\begin{figure}
\resizebox{0.50\textwidth}{!}{%
  \includegraphics{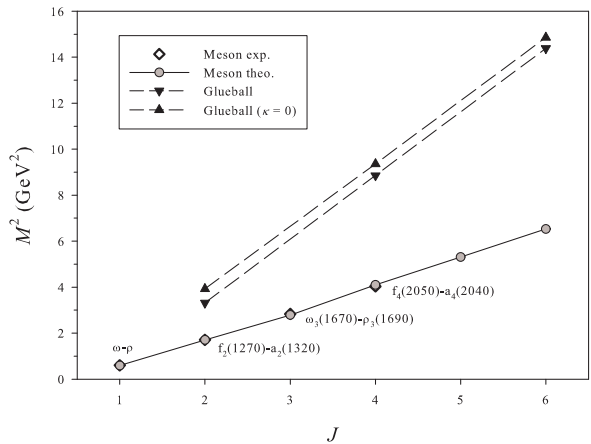}
}
\caption{Regge trajectories for light mesons (average between $I=0$
and $I=1$) and glueballs (with or without Coulomb interaction). These
states are characterized by $N=0$, $S$ maximal, and $J=L+S$.}
\label{fig:gg}
\end{figure}

In this paper, we assume that the string tension between two colour
sources is controlled by the Casimir scaling. So, contrary to what is
done in Ref.~\cite{abre05}, we assume that the ratio between the string
tension in a glueball and the string tension in a meson is given by the
ratio of the Casimir operators. In a two gluon system, we have then
$\sigma=9\, a/4$, and the Regge slope is increased by the same factor.
Moreover, following the theory of Simonov \cite{kaid00,simo05}, there is
no self-energy contribution coming from gluon. Let us note that, in the
same framework \cite{guba94,morg99}, the
constituent mass of a particle is directly proportional to
$\sqrt{\sigma}$ when its current mass $m$ is vanishing. If we compute
$\mu_g$
from a two-gluon glueball and $\mu_n$ from a light meson, we have
$\mu_g/\mu_n=3/2$. A gluon cannot decay spontaneously into a $n\bar n$
pair.

At the lowest order, the colour Coulomb factor in such a system is given
by the formula~(\ref{kappa}), that is to say $\kappa=3$. Nevertheless,
in some model, it is found that the Coulomb force in glueballs could be
damped \cite{corn83} or strongly reduced \cite{kaid00}. So, we have
considered both cases $\kappa=0$ and $\kappa=3$ to study this possible
effect. As our value of $\alpha_S$ is
small, the two spectra do not differ strongly.

The colour wave function of two gluons is symmetric and the spin wave
function considered here is symmetric since $S=2$. As gluons are bosons,
the spatial wave function is also symmetric and $L$ is
even. The first three states are computed with the six lowest light
mesons of the first Regge trajectory. Results are given in
Table~\ref{tab:m} and in Fig.~\ref{fig:gg}. When $\kappa=3$, the
trajectory is
\begin{equation}
\label{ggtraj}
J=0.36 M^2 + 0.80.
\end{equation}
When $\kappa=0$, the slope is not changed and the intercept becomes
0.57. This results is somewhat different from the results of
Refs.~\cite{abre05,donn98} because the smaller value of our string
tension $a$. But our glueball masses are compatible with
those of the model III.A of Ref.~\cite{brau05}. Moreover, the lightest
state has a mass compatible with possible experimental candidates
\cite{zou99,bugg00}.

\begin{table}
\caption{Predicted masses in GeV for $n\bar n$ (average between $I=0$
and $I=1$), $c\bar n$, $c\bar c$, glueballs, and hybrid gluelumps. These
states are characterized by $N=0$, $S$ maximal, and $J=L+S$. The
asterisk indicates that the states is calculated with $\kappa=0$.}
\label{tab:m}
\begin{tabular}{lrrrrrr}
\hline\noalign{\smallskip}
$J$ & 1 & 2 & 3 & 4 & 5 & 6 \\
\noalign{\smallskip}\hline\noalign{\smallskip}
$n\bar n$ & 0.767 & 1.306 & 1.670 & 2.024 & 2.304 & 2.555 \\
Glueball &  & 1.820 & & 2.976 & & 3.793 \\
Glueball* &  & 1.981 & & 3.059 & & 3.855 \\
$c\bar n$ & 2.033 & 2.436 & 2.747 & 3.010 & & \\
$c\bar c$ & 3.146 & 3.449 & 3.693 & 3.908 & & \\
Hybrid & & 3.820 & 4.343 & 4.749 & & \\
Hybrid* & & 4.010 & 4.464 & 4.884 & & \\
\noalign{\smallskip}\hline
\end{tabular}
\end{table}

\section{Hybrids}
\label{sec:hyb}

With formula~(\ref{kappa}), we find $\kappa=3/2$ for a gluon-quark pair
inside the hybrid. As the quark and the antiquark are assumed to be very
close, the total colour factor for the Coulomb-like force between the
gluon and the point-like $c\bar c$ pair is $2\times 3/2=3$, which is the
same as into a two gluon glueball. As in the previous case, we have
considered both $\kappa=0$ and $\kappa=3$.

The situation is different for the confinement which is actually a
three-body
force. Let us consider the general case of a triangle formed by three
colour sources named
$A$, $B$, $C$, and a point $P$ inside this triangle, where the three
flux tubes, with lengths $|AP|$, $|BP|$, $|CP|$, generated by these
sources meet \cite{silv04}. The triangular inequalities imply
$|AB| + |AC| < 2|AP| + |BP| + |CP|$. Let us assume that the energy of
these flux tubes are respectively $\sigma_A |AP|$, $\sigma |BP|$, and
$\sigma |CP|$, with $\sigma_A = k\,\sigma$. The total energy of the flux
tubes is then $\sigma (k |AP| + |BP| + |CP|)$. If $k>2$, the above
inequality shows that it is energetically favourable for the junction
point $P$ to be on the source $A$. The potential energy of the system is
then $\sigma (|AB| + |AC|)$. In the case of a hybrid gluelump, the gluon
is at position $A$ with $k=9/4$ and the quarks occupy positions $B$ and
$C$, which are merged here. So the confining energy is equal to two
times the confining energy in a meson, and we take $\sigma=2$.

To compute the mass of the hybrid, we must fix the mass of the $c\bar c$
pair. In our model, there is no direct confining interaction between the
quark and the antiquark, but a small repulsive Coulomb force exists
since formula~(\ref{kappa}) implies that $\kappa=-1/6$ for a
quark-antiquark pair in a colour octet. The $c$ and $\bar c$ quarks does
obviously not really occupy the same position. If we assume a typical
separation around the $J/\psi$ radius, the contribution of the Coulomb
interaction is around 10-20~MeV. For heavy quark, the constituent mass
$\mu$ is around the current mass $m$. For a $c\bar c$, the contribution
of the self-energy can be estimated around $-$40~MeV.
So, if we neglect the kinetic energy of the quarks assumed fixed at the
centre of mass, we can take the point-like $c\bar c$ with a mass equal
to $2\ m_c$. We estimate that this approximation can generate an error
on hybrid glueball around 100~MeV.

The masses of hybrids are given up to $J=4$ with the masses of $c\bar n$
and $c\bar c$ mesons (see Table~\ref{tab:m} and Fig.~\ref{fig:ccg}). The
hybrid masses found are significantly lower than those obtained in
Ref.~\cite{abre05}, again because the smaller value of our string
tension $a$. If the Coulomb interaction is cancelled, the masses
increase by about 150~MeV.

\section{Concluding remarks}
\label{sec:conclu}

We compute the masses of the lightest glueballs and hybrid gluelumps
(gluon with a point-like $c\bar c$ pair) within the framework of a
simple effective model of QCD derived from the rotating string model
\cite{guba94,morg99}. The corresponding semirelativistic Hamiltonian is
dominated by a linear confinement, supplemented by a one gluon exchange
interaction and a contribution from self-energy for the quarks only. It
is assumed that the Casimir scaling hypothesis is well verified, and
that two gluons are attached by an adjoint string in a glueball, while
the gluon and the colour octet $c\bar c$ pair are attached by two
fundamental strings in a hybrid gluelump. Values of effective string
tension $\sigma$ and colour factor $\kappa$ for the different hadrons
considered here are gathered in Table~\ref{tab:eff}.

\begin{table}
\caption{Values of effective string tension $\sigma$ and colour factor
$\kappa$ for different hadrons.}
\label{tab:eff}
\begin{tabular}{lrr}
\hline\noalign{\smallskip}
Hadron & $\sigma$ & $\kappa$ \\
\noalign{\smallskip}\hline\noalign{\smallskip}
Meson & $a$ & 4/3 \\
Glueball & $9 a/4$ & 3 \\
Hybrid & $2 a$ & 3 \\
\noalign{\smallskip}\hline
\end{tabular}
\end{table}

In order to minimize the contributions of the spin dependent
interactions, only the masses of states with maximal spin have been
calculated. This strongly constraints the values of our three free
parameters: the mass of the charm quark, the strong coupling constant
and the string tension in a meson. With these parameters, we find
glueballs and hybrid gluelumps with masses relatively small
\cite{abre05}. The contribution of the Coulomb interaction is about
100-200~MeV, depending on the particular hadron. The ground state
glueballs lie on a Regge trajectory and the lightest $2^{++}$ state has
a mass compatible with some experimental candidates \cite{zou99,bugg00}.
Despite the simplicity of the model, these predictions can serve as a
guide for experimental searches of exotic hadrons.


\end{document}